\documentclass{ws-ijmpcs}

\begin{document}

\markboth{L. Santos \& W. Zhao}
{Statical properties of CMB B-mode in partial sky}

%
\catchline{}{}{}{}{}
%

\title{Statical properties of CMB B-mode polarisation in a partial sky analysis}

\author{Larissa Santos}

\address{CAS Key Laboratory for Researches in Galaxies and Cosmology, Department of Astronomy, University of Science and Technology of China, Chinese Academy of Sciences, Hefei, Anhui 230026, China\\
larissa@ustc.edu.cn}

\author{Wen Zhao}

\address{CAS Key Laboratory for Researches in Galaxies and Cosmology, Department of Astronomy, University of Science and Technology of China, Chinese Academy of Sciences, Hefei, Anhui 230026, China\\
wzhao7@ustc.edu.cn}

\maketitle

\begin{history}
\received{Day Month Year}
\revised{Day Month Year}
\published{Day Month Year}
\end{history}

\begin{abstract}
Measuring the imprint of primordial gravitational waves in the cosmic microwave background (CMB) polarisation field is one of the main goals in modern cosmology. However, the so called $B$-mode polarisation can be generated by different sources besides the primary one predicted by inflationary theories, known as secondary $B$-mode signal. Among them, CMB lensing and astrophysical foregrounds play an important role. Moreover, a partial sky analysis leads to a leakage between $E$-modes and $B$-modes. In this article, we use the well known Minkowski functionals (MF) statistics to study the significance of this leakage in the CMB lensing $B$-mode signal. We find that the MF can detect the $E$-to-$B$ leakage contamination, thus it should not be neglected in future CMB data analysis.

\end{abstract}

\ccode{PACS numbers:95.85.Sz, 98.70.Vc, 98.80.Cq}

\section{Introduction}	

The cosmic microwave background (CMB) radiation is revealing the physics of the early universe since it was measured by the first time in 1965.  According to the standard cosmological model, the full CMB information can be extracted by analysing the statistical properties of both temperature,  $T(\hat{\gamma})$, and polarisation fields. The latter are described by the stocks parameters, $Q(\hat{\gamma})$ and $U(\hat{\gamma})$, usually decomposed into the curl-free ($E$-mode) and divergence free ($B$-mode) components\cite{zaldarriaga-b-mode,kamionkowski-b-mode}. Together, these are powerful observables to understand the evolution of the cosmos and to probe our cosmological models. Inflationary theories predict the existence of a gravitational wave stochastic background that would generate an imprint in the CMB polarisation field, known as primordial $B$-mode signal\cite{zaldarriaga-b-mode,kamionkowski-b-mode,gw-b-mode}. In order to probe inflation, the measurement of the primordial $B$-mode signal is then the main target of future CMB experiments. One problem arises: to distinguish the primordial $B$-mode signal from the secondary $B$-mode signals generated by CMB lensing\cite{lewis-review}, astrophysical foregrounds\cite{foregrounds,planck-foregrounds}, and the leakage between $E$- and $B$-modes\cite{ebmixture1,ebmixture2,ebmixture3}. Our ability to decompose the secondary CMB $B$-mode polarisation signal in a partial sky coverage is of great importance in order to resolve the primordial signal.

We already know that in the highly non-Gaussian lensed $B$-map the astrophysical residuals can be detected by the MF if more than $0.4\%$ of foreground radiation is still present on the map\cite{larissa}. Here, we do not consider astrophysical foregrounds. Instead, we focus in characterizing the imprint of the $E$-$B$ mixture in the lensed $B$-mode using the well-known Minkowski functionals (MF)\cite{schmalzing1998}.

\section{$E$- and $B$-mode decomposition in partial sky}
\label{$E$- and $B$-mode decomposition in partial sky}

The Stokes parameters $Q$ and $U$, combined into a spin-(2) and spin-(-2) fields $P_{\pm}(\hat{n})=Q(\hat{n})\pm iU(\hat{n})$, describe completely the linearly polarised CMB field. For full sky, the spin fields can be expanded over spin-weight harmonic functions basis\cite{seljak1996}, such that $P_{\pm}(\hat{n})=\sum_{\ell m} a_{\pm2,\ell m}~_{\pm 2}Y_{\ell m}(\hat{n})$. The $B$-mode field can be written in terms of the coefficients $a_{\pm 2,\ell m}$, being the polarisation map,  $B(\hat{n})$, defined in terms of spherical harmonics:

\begin{eqnarray}
\label{eq_elm}
B_{\ell m}\equiv -\frac{1}{2i}[a_{2,\ell m}-a_{-2,\ell m}], ~~~~~
B(\hat{n})\equiv \sum_{\ell m}B_{\ell m} Y_{\ell m}(\hat{n}).
\end{eqnarray}

Considering that the polarisation field is measured only in a fraction of the sky, we must define a new field $\mathcal{B}$ in order to derive the $B$-mode coefficients\cite{zaldarriaga-seljak}.

\begin{eqnarray}
\label{pseudo_B}
\mathcal{B}(\hat{n}) = -\frac{1}{2i}[\bar{\eth}\bar{\eth}P_{+}(\hat{n}) - \eth \eth P_{-}(\hat{n})],~~~~~~
\label{pseudo_B_harm}
\mathcal{B}(\hat{n}) \equiv \sum_{\ell,m}\mathcal{B}_{\ell m}Y_{\ell m}(\hat{n}),
\end{eqnarray}

where $\eth (\bar{\eth})$ corresponds to the spin-raising (lowering) operator, and $\mathcal{B}_{\ell m} =  \int  \mathcal{B}(\hat{n})Y_{\ell m}^*(\hat{n})d\hat{n}$. These pseudo multipoles are related to the regular $B_{\ell m}$ by  $\mathcal{B}_{\ell m}=N_{\ell,2}B_{\ell m}$, being $N_{\ell ,s}=\sqrt {{(\ell +s)!}/{(\ell -s)!}}$.

 We now introduce a window function $W$\cite{efstthiou2004} that leads to $\tilde{\mathcal B}_{\ell m} = \int d\hat{n}W(\hat{n}) \mathcal{B}(\hat{n})Y_{\ell m}^*(\hat{n})$. The pure $B$-mode pseudo multipoles can be equivalently defined as

\begin{align}
\label{pureb}
\mathcal{B}_{lm}^{\text{pure}}\equiv-\frac{1}{2i}\int d\hat{n}\left\{P_{+}(\hat n)\left[\bar {\eth} \bar{\eth}\left(W(\hat{n})Y_{\ell m}(\hat{n})\right)\right]^\ast-P_{-}(\hat n)\left[ {\eth} {\eth}\left(W(\hat{n})Y_{\ell m}(\hat{n})\right)\right]^\ast \right\},
\end{align}

In the above definition, the window function and its first derivative, $\partial W$, must vanish at the observed patch boundaries. With this in mind, we chose the Gaussian smoothing method\cite{kim2011,Wang} to smooth the edges of $W$,

\begin{equation}
\label{WF}
W =\left \{ \begin{array} {ll}
\int_{-\infty}^{\delta_i - \frac{\delta_c}{2}} \frac{1}{\sqrt{2\pi\sigma^2} } \exp\left(-\frac{x^2}{2\sigma^2}\right)dx=\frac{1}{2} + \frac{1}{2} \text{erf} \left(\frac{\delta_i - \frac{\delta_c}{2}}{\sqrt 2 \sigma}\right) \quad \delta_i < \delta_c,\\
1 \quad  \quad \quad \quad \quad \quad \quad \quad \quad \quad \quad \quad \quad \quad \quad \quad \quad \quad  \quad \quad \quad  \quad \delta_i >\delta_c
\end{array} \right.
\end{equation}

Where $\delta_i$ the smallest angular distance between the $i$-th observed pixel and the boundary of the mask. $\delta_c$ is an adjustable parameter referred as the apodization length. Throughout this article, we use the method developed by Smith and Zaldarriaga (hereafter, SZ)\cite{ebmixture3} to extract the $E$ and $B$ signals from partial $Q$ and $U$ sky.

\section{Minkowski functionals}
\label{MF}

The MF describe the morphological properties of convex, compact sets in an $n$-dimensional space. On a 2-dimensional CMB field defined on the sphere,  $\mathcal{S}^2$, the morphological properties of the data can be characterized as a linear combination of three MF. We can define then the excursion set, or connected region, $Q_{\nu}=\{x\in \mathcal{S}^2|u(x)>\nu\sigma\}, $ of a scalar field, $u$, of zero mean and of variance $\sigma^2$, for a certain threshold, $\nu$, such that $u(x)/\sigma> \nu$. Its boundary is defined as $\partial Q_{\nu}=\{x\in \mathcal{S}^2|u(x)=\nu\sigma\}$. For the CMB field, we then have the area $v_0(\nu)$, the contour length,  $v_1(\nu)$ and the integrated geodetic curvature, $v_2(\nu)$,  as \cite{schmalzing1998}:

 \begin{equation}
  \label{MFs}
 v_0(\nu)=\int_{Q_{\nu}}\frac{da}{4\pi},~v_1=\int_{\partial Q_{\nu}}\frac{d l}{16\pi},~v_2=\int_{\partial Q_{\nu}}\frac{\kappa dl}{8\pi^2},
 \end{equation}

where $da$ and $dl$ are the surface element of $\mathcal{S}^2$ and the line element along $\partial Q_{\nu}$, respectively, and $k$ is the geodetic curvature. The MF can be numerically calculated for a given pixelized map in a simple way\cite{schmalzing1998,lim2012}. Here, we  study the statistical properties of the leakage from $E$- to $B$-modes due to a partial sky analysis. The algorithm for calculating the MF was developed by Gay {\it et. al} (2012) and Ducout {\it et.al} (2013)\cite{gay 2012,ducout2012}.

\section{Methodology}
\label{Methodology}

In the analysis throughout this article, we used simulated CMB maps generated by the Lenspix software. Our simulations had 500 full sky $Q$ and $U$ lensed maps with cosmological parameters $h^2\omega_b=0.0223$,  $h^2\omega_b=0.1188$, $h=0.673$, $A_s=2.1*10^{-9}$, $n_s=0.9667$, $r=0$, $FWHM=30'$ and $nside=1024$.  First, we obtained the $B$-maps directly from the full sky $Q$ and $U$ maps, hereafter called the {\bf ideal case}. Moreover, we numerically obtained a second set of $B$-maps, now generated using the SZ E/B decomposition method\cite{ferte2013} where the Galactic region was masked using the smoothed apodized window function derived from the Planck UT78 polarization mask (see Eq.(\ref{WF})). These final $B$-maps are, from now on, called the {\bf real case}.

In order to remove the contribution from multipoles dominated by noise, the calculation of the MF  requires that we smooth the maps before analysing them. Thus, we smooth each final $B$-map for both ideal and real cases using a Gaussian filter with 6 different smoothing scales (to extract all the available statistical information),  $\theta_s=10', 20', 30', 40', 50', 60'$, generating 6 sets of 500 maps for each considered case. It is important to point out that the information extracted of the CMB is dominant in a different multipole range for each smoothing scale\cite{hikage2006,hikage2008,ducout2012}. We statistically analyse these final $B$-maps by means of the MF, with binning range of the threshold $\nu$ set from $-3$ to $3$ with 25 equally spaced bins.

Finally, we compare the real and ideal cases to look for the $E$-to-$B$ contamination imprint present only in the real case. Note that, for the ideal case, we also applied the same smoothed apodized window function derived from the Planck UT78 polarization mask to calculate the MF to ensure we are comparing the same regions in the sky.

\section{Results}
\label{Results}

\begin{figure}[ht]
\centerline{\includegraphics[width=16cm,height=10cm]{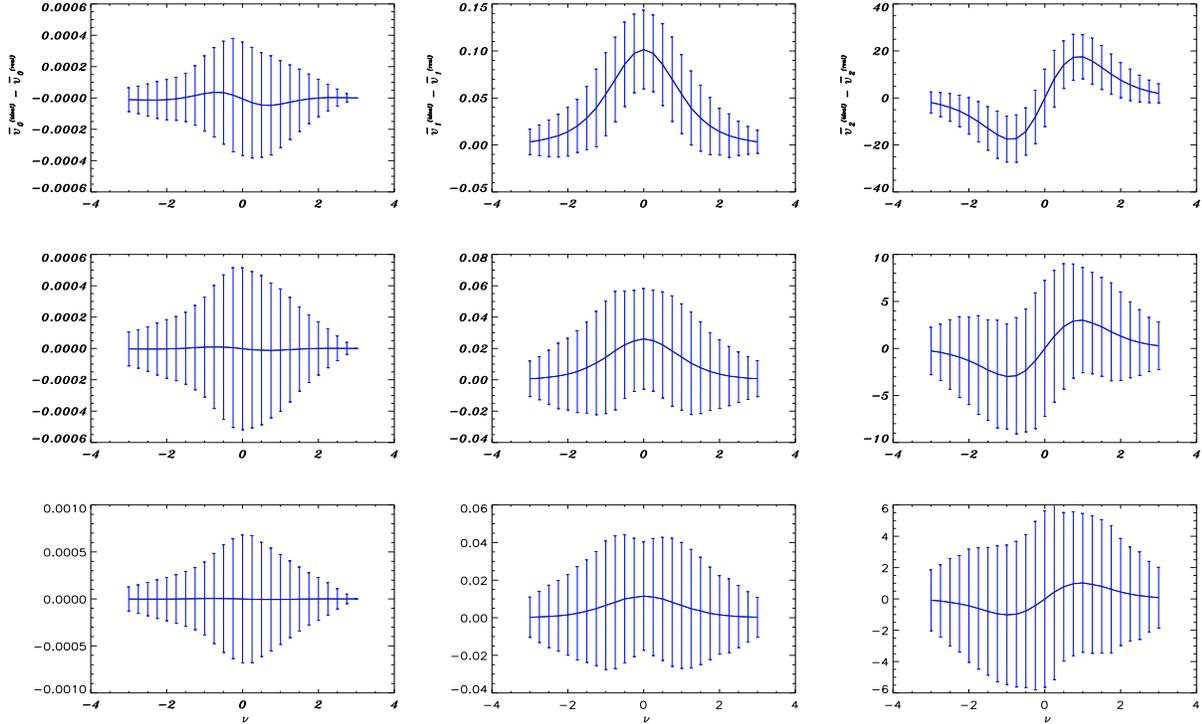}}
\caption{The difference between the mean values of real and ideal case for the MF considering $r=0$ over 500 simulations. From top to bottom: $\theta_s=10', 40', 60'$, respectively. From left to right: the first, second and third MF, respectively.}
\label{MF_r0}
\end{figure}

The use of masks to avoid Galactic foregrounds, necessary even when the data is collected by satellite surveys, leads to the so called $E$-to-$B$ leakage. Together with other secondary $B$-mode signals (for example, CMB lensing and astrophysical foregrounds), it can mimic the primordial signal.  We look for an imprint of the $E$-to-$B$ leakage, in order to distinguish it from the primordial signal, by employing the MF to our final $B$-maps.  We found this imprint of the leakage in the difference between the real case and the ideal case, as it can be seen in Fig. \ref{MF_r0} for $v_0, v_1, v_2$.  As expected due to the loss of information, by increasing $\theta_s$  the significance of the leakage becomes smaller, being the signal more evident at lower smoothing scales.  To quantify this result, we calculate the $\chi^2$ statistics defines as

 \begin{equation}
  \label{chi}
\chi^2 = \sum_{aa'} \left[\bar{v}_a^{ideal} - \langle v_a^{real} \rangle \right] C^{-1}_{aa'} \left[\bar{v}_{a'} ^{ideal} - \langle v_{a'}^{real} \rangle \right],
 \end{equation}

where $\langle \bar{v}_a^{real} \rangle $ is the model under test.  For each smoothing factor, $\theta_s$,  $a$ and $a'$ denote the binning number of the threshold value $\nu$ and the different kinds of MF.  The covariance matrix is estimated from the average under 500 simulations $C_{aa'} \equiv  \frac{1}{499} \sum_{k=1}^{500} \left[\left(v_a^{k,real} -  \bar{v}_a^{real}  \right)\left(v_{a'}^{k,real} -  \bar{v}_{a'}^{real}  \right)\right]$. The results are shown in Table \ref{T1}.

\begin{table}
\tbl{Significance of the $E$-to-$B$ leakage in terms of the $\chi^2$ statistics.}
{\begin{tabular}{c c c c c c }
\hline\hline
\toprule $\theta_s=10'$  & $\theta_s=20'$  &  $\theta_s=30'$  &  $\theta_s=40'$ & $\theta_s=50'$ & $\theta_s=60'$ \\
\colrule
  18.93 & 9.53  & 4.27 &1.62 & 0.71 & 0.34\\
\hline
\botrule
\end{tabular}
\label{T1}}
\end{table}

 The total $\chi^2_T$ is then obtained considering the combination of every smoothing scale, so that  $a$ and $a'$ also denote $\theta_s$ in the definition for the $\chi^2$ stated in Eq. (\ref{chi}). We found that $\chi^2_T= 220.23$, making it clear that the MF for different smoothing scales are very correlated since the leakage is not a stochastic noise. Therefore, it is important to emphasize that even though the leakage seems unimportant for individual smoothing scales, it is indeed relevant when they are all combined.

\section{Discussion and conclusions}
\label{Conclusions}

The CMB primordial $B$-mode signal is the main target of future cosmological experiments since it can give important information about the physics of the early universe. However, this primordial signal can be contaminated by secondary ones generated by astrophysical foregrounds, CMB lensing or caused by a leakage from $E$-to $B$ modes due to a data analysis in an incomplete sky. Here, we analysed the significance of the leakage in the CMB $B$-map simulations by means of the MF statistics. In this first analysis we did not consider any contribution from primordial gravitational waves signal, but we included the CMB lensing effect on the simulations which also generates $B$-mode polarisation.  We found that the leakage contribution is not negligible when all the available information stored in the MF is considered, i.e., combining all the smoothing scales. In order to avoid misinterpretation of the data, the E-to-B leakage must be taken into account when analysing the CMB data in an incomplete sky survey. In a future analysis, in order to corroborate this result, we intend to include the primordial the gravitational wave signal as a non zero tensor-to-scalar ratio, and to use different statistics.

\section*{Acknowledgments}
We acknowledge the use of the Planck Legacy Archive (PLA). Our data analysis made the use of HEALPix \cite{healpix}, CAMB \cite{camb} and LensPix \cite{lenspix}. This work is supported by NSFC No. J1310021, 11603020, 11633001, 11173021, 11322324, 11653002, 11421303, project of Knowledge Innovation Program of Chinese Academy of Science and the Fundamental Research Funds for the Central Universities.



\end{document}